# On Exponential-Time Completeness of the Circularity Problem for Attribute Grammars

Pei-Chi Wu

Department of Computer Science and Information Engineering
National Penghu Institute of Technology
Penghu, Taiwan, R.O.C.
E-mail: pcwu@npit.edu.tw

## Abstract

Attribute grammars (AGs) are a formal technique for defining semantics of programming languages. Existing complexity proofs on the circularity problem of AGs are based on automata theory, such as writing pushdown acceptor and alternating Turing machines. They reduced the acceptance problems of above automata, which are exponential-time (EXPTIME) complete, to the AG circularity problem. These proofs thus show that the circularity problem is EXPTIME-*hard*, at least as hard as the most difficult problems in EXPTIME. However, none has given a proof for the EXPTIME-completeness of the problem. This paper first presents an alternating Turing machine for the circularity problem. The alternating Turing machine requires polynomial space. Thus, the circularity problem is in EXPTIME and is then EXPTIME-complete.

# 1. INTRODUCTION

Attribute grammars (AGs) [8] are a formal technique for defining semantics of programming languages. There are many AG systems (e.g., Eli [4] and Synthesizer Generator [10]) developed to assist the development of "language processors", such as compilers, semantic checkers, language-based editors, etc. These AG systems process language specifications written in AGs and generate corresponding language processors.

An AG is *circular* if there is a cycle in (attribute) dependency graphs. A circular dependency is thought to be a specification error and should be detected by AG systems. Knuth [8] presented an exponential-space algorithm for the circularity problem. The intrinsically exponential complexity of the circularity problem for AGs was first proved by Jazayeri et al. [6], who reduced the acceptance problem of writing pushdown acceptors [10] to the circularity problem. Jazayeri [7] (and the correction by Dill [3]) tried to provide a simpler construction of AGs by reducing the acceptance problem of space-bounded alternating Turing machines [1] to the circularity problem. The acceptance problem of (polynomial) space-bounded writing pushdown acceptors and alternating Turing machines are exponential-time (EXPTIME) complete. Both reductions show that the circularity problem for AGs is EXPTIME-*hard*, at least as hard as the most difficult problems in EXPTIME. However, none has given a proof for the EXPTIME-completeness of the problem.

To show the circularity problem for AGs to be EXPTIME-complete, we need an algorithm that requires exponential time but not actually exponential space. The existing circularity test algorithms [2, 6, 8, 9] all require exponential space in the worst cases. In fact, it is almost unlikely to find an algorithm for the circularity problem that takes only polynomial space. If there is one, the circularity problem is in polynomial space (PSPACE) and PSPACE $\supseteq$ EXPTIME. Since PSPACE $\subseteq$ EXPTIME, we would get

PSPACE = EXPTIME, which is still unknown in complexity theory (cf. e.g., Chandra et al. [1]).

This paper presents an alternating Turing machine for the circularity problem. The alternating Turing machine requires polynomial space. Thus, the circularity problem is in EXPTIME and is then EXPTIME-complete.

## 2. THE ALTERNATING ALGORITHM

In an alternating Turing machine [1, 7], there are two main kinds of states: *universal* and *existential*. An existential state leads to an acceptance of the input, if *one* of the alternatives (possible next moves) leads to an accepting state. A universal state leads to an acceptance of the input, if *all* of the next moves lead to an accepting state.

Algorithm *CircularityTest* is an EXPTIME algorithm represented using an alternating Turing machine. Since complete development of the algorithm using "primitive" operations, e.g., moving heads and writing tape symbols, in Turing machines is very complex, the algorithm is sketched using higher-level operations, borrowed from programming languages. Typical sequential code fragments are written directly in Pascal style pseudo codes, which can easily be converted into alternating Turing machine operations. The "guess" actions in states $q_1$ and $q_3$ can be programmed using a series of existential states, each of which randomly chooses or does not choose an arc in a dependency subgraph. To explore the parallelism in alternating Turing machines, a FORK statement creates a number of parallel tasks (possible next moves). Each task is given its next state and a number of input parameters, which are placed on the working tape. Each state is marked as universal or existential. A task ends with either success or failure: There is no task return, and a run-time stack is not needed.

**Algorithm** *CircularityTest*.

*Input.* An AG *ag*. *P* denotes its production rules, *N* denotes its nontermnal symbols, *Inh*(*X*) (*Syn*(*X*)) denotes the inherited (synthesized) attributes of symbol *X*, and *D*(*p*) denotes the dependency graph of a production *p*.

*Output.* Whether *ag* is circular or not: *ag* is circular if *success*, non-circular if *failure*.

*Method.*

   State $q_0$: (existential)

      **FORK** $q_1(p)$, $p \in P$;

   State $q_1(p)$: (existential)

      **IF** $p$: $X_0 \rightarrow \varepsilon$

      **THEN** *failure*; {* There is no cycle in the dependency subgraph of $X_0$ *}

      **ELSE**

         Let $p$: $X_0 \rightarrow X_1 \ldots X_k$;

         Guess an array of dependency subgraphs $[d_1, \ldots, d_k]$ for $X_i$,

            where $d_i \subseteq Inh(X_i) \times Syn(X_i)$, $i > 0$;

         Compose dependency subgraphs in $[d_1, \ldots, d_k]$ with $D(p)$;

         **IF** a (potential) cycle is found in the resulting dependency graph

         **THEN FORK** $q_2(X_i, d_i)$; {* verify each of dependency subgraph $d_i$ of $X_i$ *}

         **ELSE** *failure*;

   State $q_2(X; d)$: (universal)

      **FORK** $q_3(p, d)$, $\forall\, p \in P$, $p$: $X \rightarrow \ldots$ ; {* succeeds if one production *p* satisfies *}

   State $q_3(p; d)$: (existential)

      **IF** $p$: $X_0 \rightarrow \varepsilon$ and $d = D(p)$

      **THEN** *success*;

      **ELSE**

         Let $p$: $X_0 \rightarrow X_1 \ldots X_k$;

         Guess an array of dependency subgraphs $[d_1, \ldots, d_k]$ for $X_i$,

            where $d_i \subseteq Inh(X_i) \times Syn(X_i)$, $i > 0$;

    Compose dependency subgraphs in $[d_1, ..., d_k]$ with $D(p)$;

    **IF** $d$ equals the resulting dependency subgraph induced at $X_0$

    **THEN FORK** $q_2(X_i, d_i)$, $i > 0$; *{\* verify each of dependency subgraph $d_i$ of $X_i$ \*}*

    **ELSE** *failure*;

**End of Algorithm.**

State $q_0$ is the initial state. The code *success* enters an accepting state, and the code *failure* enters a rejecting state. All states except $q_2$ are existential. State $q_2$ is universal and is called from state $q_1$ and $q_3$ to verify that all dependency subgraphs that cause a cycle can be generated. The space needed in the algorithm is proportional to the size of the input AG, i.e., $O(n)$: The input tape contains the input AG, and the working tape contains the area for placing parameters, which is limited by the size of largest parameters. Other operations such as composing dependency graphs and detecting cycles can also be programmed using $O(n)$ space.

The correctness of the algorithm can easily be verified. The algorithm is a reverse version of the original circularity test algorithm [8]: First, guess an array of possible dependency subgraphs. Second, if there exists a cycle in the composition of these subgraphs, make sure that all the dependency subgraphs can be generated. The second step recursively calls itself (the calling sequence $q_2$-$q_3$-$q_2$-$q_3$-...). This is similar to the step that collects dependency graphs repeatedly in the original algorithm.

The presented alternating Turing machine needs only polynomial space; however it cannot be converted into a space-efficient algorithm in computers (or deterministic Turing machines), because alternating Turing machines are very powerful and have almost unlimited parallelism and memory space to create and execute new parallel tasks. A straightforward simulation of the presented alternating Turing machine needs exponential time and space, which is not better than the complexity of the existing circularity test algorithms. In addition, this algorithm may be even slower than original algorithms, because this algorithm minimizes the space needed for each task but does not care how much redundant work is performed by the numerous parallel tasks.

Because the algorithm contains a universal state $q_2$, it cannot be ported to a nondeterministic Turing machine. This is important. As already known, PSPACE = NPSPACE. If there exists a nondeterministic Turing machine for the circularity problem, we could get PSPACE = EXPTIME, a contradiction.

## 3. CONCLUSIONS

We have presented an alternating Turing machine for the circularity problem. The alternating Turing machine requires polynomial space. Thus, the circularity problem is in EXPTIME and is then EXPTIME-complete.

**REFERENCES**

1. Chandra, A. K., Kozen, D. C., Stockmeyer, L. J., "Alternation," *Journal of the ACM*, Vol. 28, No. 1, Jan 1981, pp. 114-133.
2. Deransart, P., Jourdan, M., Lorho, B., "Speeding up Circularity Tests for Attribute Grammars," *Acta Informatica*, Vol. 21, pp.375-391, 1984.
3. Dill, J.M., "A Counterexample for 'A Simpler Construction for Showing the Intrinsically Exponential Complexity of the Circularity Problem for Attribute Grammars," *Journal of the ACM*, Vol.36, No.1, Jan. 1989, pp.92-96.
4. Gray, R.W., Heuring, V.P., Levi, S.P., Sloane, A.W., Waite, W.M., "Eli: A Complete, Flexible Compiler Construction System," *Communications of the ACM*, Vol.35, No.2, Feb. 1992, pp.121-131.
5. Hopcroft, J. E. and Ullman, J. D., *Introduction to Automata Theory, Languages, and Computation*, Addison-Wesley, 1979.
6. Jazayeri, M., Ogden, W.F., Rounds, W.C., "The Intrinsically Exponential Complexity of the Circularity Problem for Attribute Grammars," *Communications of*

*ACM*, Vol. 18, No. 12, Dec. 1975, pp. 697-706.

7. Jazayeri, M., "A Simpler Construction for Showing the Intrinsically Exponential Complexity of the Circularity Problem for Attribute Grammars," *Journal of the ACM*, Vol.28, No.4, Oct. 1981, pp.715-720.

8. Knuth, D. E., "Semantics of Context-Free Languages," *Mathematical Systems Theory*, Vol. 2, No. 2, 1968, pp. 127-145. Correction: *Mathematical Systems Theory*, Vol. 5, No. 1, 1971, pp. 95-96.

9. Räihä, K.-J., Saarinen, M., "Testing Attribute Grammars for Circularity," *Acta Informatica*, Vol. 17, pp.185-192, 1982.

10. Reps, T. and Teitelbaum, T., *The Synthesizer Generator: A System for Constructing Language-Based Editors*, Springer-Verlag, Newyork, 1989.